\documentclass[preprint,aps,showpacs,pre,amsmath]{revtex4-2}

\usepackage{amssymb}
\usepackage{graphicx}
\usepackage{dsfont}
\usepackage{color,soul}
\usepackage{siunitx}
\usepackage{babel,blindtext}
\usepackage[labelformat=empty]{caption}
\usepackage{tabularx}
\usepackage{multirow}
\usepackage{xcolor}
\usepackage[normalem]{ulem}
\usepackage{soul}

\begin{document}

\title{First-principles study of the structural and electronic properties of BN-ring doped graphene}

\author{L. Caputo, V.-H. Nguyen and J.-C. Charlier}
\affiliation{Institute of Condensed Matter and Nanosciences, Universit\'{e} catholique de Louvain (UCLouvain), Chemin des \'etoiles 8,  B-1348 Louvain-la-Neuve, Belgium}

\begin{abstract}
Since advanced Silicon-based device components are moderately chemically tunable, doped graphene has emerged as a promising candidate to replace this semiconducting material in flexible miniaturized electronic devices. Indeed, heteroatom co-doping (i.e. with boron and/or nitrogen) is an appealing strategy to tune both its structural and electronic properties, possibly inducing a band gap in graphene. However, presently synthesized BN-doped carbon-based materials are randomly doped, leading their electronic properties not to be reproducible.
Using first-principles techniques, the present study investigates the periodic doping of graphene with borazine-like rings in order to search for an entirely new class of BCN hybrid 2D materials exhibiting high stabilities and optimized band gaps for opto-electronic applications.
{\it Ab initio} calculations show that BN-ring doped graphene displays cohesive energies comparable with benchmark ideal periodic BCN systems (such as BC$_3$, C$_3$N$_4$, BC$_2$N) with a decreasing linear trend toward high concentrations of BN-rings. Band gaps of BN-ring doped graphene systems are calculated using many-body perturbation techniques and are found to be sensitive to the doping pattern and to be considerably larger for high concentration of BN rings exhibiting the same orientation. These predictions suggest that BN-ring doped graphene materials could be interesting candidates for the next generation of optoelectronic devices and open new opportunities for their synthesis using chemical bottom-up approaches.

\end{abstract}
\maketitle

\section{Introduction}
Nowadays, the design of new materials is a fundamental issue driven by the need to improve existing technology devices for better performance and enhanced safety. Currently, the majority of opto-electronic devices are based on doped inorganic semiconductors (i.e. Si) \cite{Si}. Their doping allows the systematic tuning of the band alignment at the interface of semiconductors, increasing its conductivity since substitutional dopant atoms donate mobile charge carriers. However, Si-based devices have numerous drawbacks, such as high production cost, stiffness, and size limitations that hinder the prospect of current technology transitioning into flexible and miniaturized devices. Moreover, they exhibit indirect band gap, low carrier mobility, and are at best only moderately tunable, which limits the possibility of band alignment. For all of the aforementioned reasons, efforts have been spent in order to search for new replacing materials \cite{Katz,Kelley}. Graphene has emerged as an excellent candidate for this role thanks to its remarkable properties, such as atomic-layer thickness, large surface area, high carrier mobility, flexibility, as well as high thermal and chemical stability \cite{Ferrari}. Indeed, its electronic properties are characterized by a band degeneracy that occurs at the two points (K, K') at the corners of the hexagonal Brillouin zone of graphene with a corresponding linear energy dispersion, resulting in the formation of Dirac cones \cite{Wallace}. This specific electronic behavior leads to a very small on-off current ratio that makes graphene unsuitable for conventional electronic applications. Essentially, exfoliated graphene exhibits no band-gap \cite{Novoselov} thus requiring post-processing to be implemented in opto-electronic devices and light-triggered applications. Different approaches have been investigated to efficiently open a band gap in graphene, such as quantum confinement within the fabrication of nanoribbons \cite{Han} or based on its specific deposition on different substrates \cite{Jung,Zhou}. However,  these methodologies do not allow the systematic and controllable tuning of the electronic properties of graphene. Consequently, fusing $sp^2$-C scaffolds with isoelectronic and isostructural BN-domains has emerged as an appealing strategy to modify both its electronic and structural properties. Indeed, BN-based materials are chemically inert and are thus used as dielectric spacer layers in van der Waals heterostructures or opto-electronic devices \cite{Britnell}. In contrast to graphene, the presence of two different atoms in the two sublattices of $h$-BN precludes the inversion symmetry, resulting in degeneracy lifting at the Dirac points in the Brillouin zone. Consequently, BN-based materials are characterized by large band gaps (around $\sim$6eV in its pristine form\cite{BN}) which strongly restrain their implementation as semiconductors in electronics.However, doping graphene with BN domains permits the opening of a band gap. In particular, the honeycomb lattice of graphene is formed by two Carbon atoms (labeled as A and B, respectively) in its primitive cell. Each A(B)-type Carbon atom is surrounded by three other B(A)-type ones thus forming two  symmetrical-triangular sublattices. The symmetry between these two sublattices leads to the gapless character of pristine graphene \cite{Castro09}. Consequently, the incorporation of Boron and Nitrogen atoms can induce an on-site energy asymmetry in these two sublattices (i.e. chiral symmetry breaking) that eventually opens a band gap at the Dirac points. Within this conceptual approach, specific doping patterns can be used to tune the electronic properties, thus inducing a wide range of band gaps in BN-doped graphene. Despite the integrity of each individual phase being maintained, allowing easier fabrication, the current state-of-the-art is still far from a targeted synthesis of BN-doped 2D materials exhibiting specific and controlled properties. In fact, the existing BN-doped carbon-based materials are not periodically doped, leading to non-reproducible properties \cite{Ci}. More specifically, exploitable materials featuring precise doping patterns of B, N and atoms are particularly difficult to achieve since atom segregation prevails \cite{Huang,Yuge} resulting from the large binding energy between boron-nitrogen and carbon-carbon atoms. In fact, to our knowledge, only one example of BN-doped covalent network featuring a regular doping pattern has been obtained so far through surface-assisted reaction \cite{Sanchez}. Consequently, an open challenge in the fabrication of functional single-layer semiconducting materials is to gain control of their structural and electronic properties in a controllable and reproducible manner. In order to achieve this goal, preliminary screening of BN doping patterns is necessary to support the development of new synthetic strategies for BNC nanomaterials. Using theoretical modelling, different parameters such as the position, the distance of the doping unit, and even the orientation can be easily tuned in order to investigate the corresponding effect on the properties of the material. In this work, state-of-the-art DFT calculations are performed on several BN-doped graphene monolayers using borazine (BN)$_3$ doping units in order to predict the corresponding structural and electronic properties of novel BNC monolayers. After evaluating the stability of these BN-ring doped graphene when considering different doping parameters, accurate band gap values are predicted using beyond DFT techniques in order to propose potential candidates to be implemented in optoelectronic nanodevices. Lastly, the disorder related to ring location and rotation is also investigated to estimate its effects on the electronic properties of BNC materials.
	
\section{Methodology}
Theoretical modelling was based on density functional theory (DFT) calculations with the projector-augmented wave method \cite{Kresse} and plane waves basis as implemented in the Vienna Ab initio Simulation Package (VASP) \cite{VASP1,VASP4}. The generalized-gradient approximation of Perdew Burke and Ernzherof \cite{PBE} has been used for the exchange-correlation density functional. Convergence threshold for total energy and forces were set to 10$^{-5}$ eV and below 0.03 eV/$\si{\angstrom}$ on each atom, respectively. All calculations have been performed using 500 eV as kinetic energy cut-off. A $k$-point sampling of 18x18x1 has been used for pristine graphene. To sample multiple BN concentrations and patterns, graphene supercells of different sizes have been constructed, with accordingly scaled and converged $k$-point samplings for every system. In every model, atomic positions of all atoms have been allowed to relax in the supercell. Graphene unit cell has been relaxed starting from experimental parameters \cite{Wickoff} using 10$\si{\angstrom}$ of vacuum in order to avoid spurious interactions along the z-direction and the obtained theoretical lattice parameters deviate $\sim$0.5$\%$ from the experimental ones. 

Regarding graphene doped with BN segregated islands, a 10x10 supercell has been considered in order to ensure sufficient separation and to avoid spurious coupling between BN islands.  The cohesive energy of each atomistic model has been calculated as follows:
\begin{equation}
    E_{coh} = \frac{-E_{BNC} + n_{C}E_{C} + n_{B}E_{B} + n_{N}E_{N}}{n_{tot}}
\end{equation}
where $E_{coh}$ is the cohesive energy, $E_{C}$, $E_{B}$ and $E_{N}$ are the total energies of isolated Carbon, Boron and Nitrogen atoms respectively while n$_{C}$, n$_{B}$ and n$_{N}$ represent the corresponding number of atoms in the atomistic model, n$_{tot}$ being the total number of the atoms that form the BNC system.\\

In order to predict accurate band gaps and to avoid their usual DFT underestimation, many-body perturbation theory calculations \cite{Hedin} employing screened Couloumb interaction were performed without self-consistency in Green's function (G$_0$W$_0$ approximation) using Quantum Espresso \cite{QE} with norm conserving pseudo-potentials \cite{PseudoDojo} and YAMBO code \cite{Yambo1}. Plasmon-pole approximation for the dielectric function was used and a truncated Coulomb potential approach was employed in the z-direction to avoid spurious interactions between periodically repeated images \cite{Rozzi}. Moreover, the Random Integration Method was also used in order to avoid numerical divergences that could be present in many-body calculations on low-dimensional systems \cite{Yambo1}. Cutoff energies for both exchange and correlation parts of the self-energy were converged for each model ranging from 50 -- 70 Ry and between 10 -- 13 Ry respectively. The number of bands used for each calculation was converged using the Bruneval-Gonze terminator \cite{BG}, which permits faster convergence with respect to the number of empty bands.

\section{Results and discussion}

\begin{figure*}[!h]
\includegraphics[width = 0.35\textwidth]{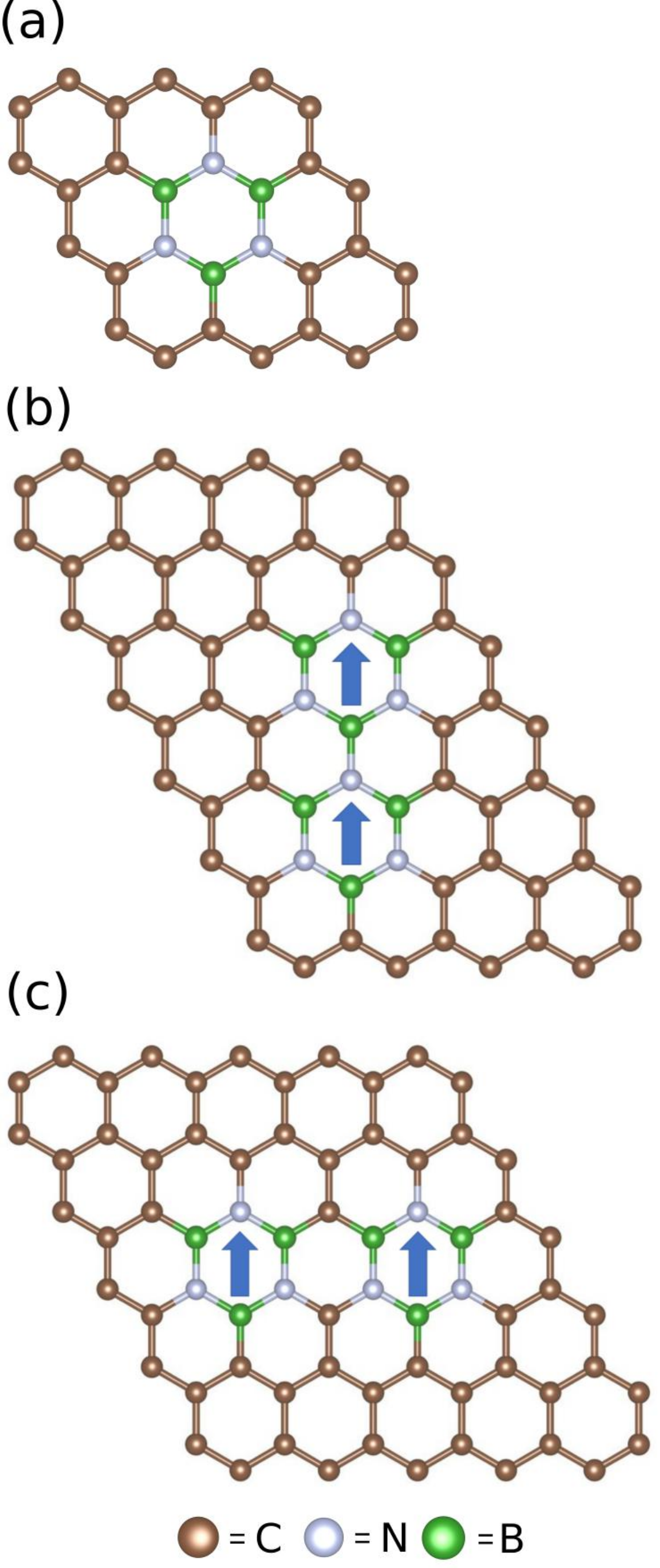}
\caption{Figure 1: Atomistic models of graphene doped with various borazine (BN)$_3$ patterns. (a) single borazine doping. (b) doping pattern formed by two (BN)$_3$ rings along the armchair direction with parallel orientation (up$\uparrow$-up$\uparrow$). (c) doping pattern formed by two (BN)$_3$ rings along the zigzag direction with parallel orientation (up$\uparrow$-up$\uparrow$). Carbon, Nitrogen and Boron atoms are represented by brown, grey and green spheres, respectively.}
\label{fig_sim1}
\end{figure*}

In order to compare the stability of novel BNC materials, the cohesive energy of graphene (7.85 eV), $h$-BN (6.99 eV), BC$_3$ (7.25 eV/atom), C$_3$N$_4$ (6.04 eV/atom), BC$_2$N (7.19 eV/atom) have been first estimated. All the atomistic models are reported in the Supplementary Information \cite{SupInf}. As expected, graphene presents the highest stability. Starting from the graphene lattice, several supercells have been built and subsequently doped with various BN-ring doping patterns, taking into account different distances, orientations and concentrations (see Fig.1). Firstly, a doping pattern formed by a single borazine (BN)$_3$ ring (Fig.1.a) is considered (referred as model 1 in the following). Afterwards, two families of doping patterns formed by two borazine rings are studied along both armchair (Fig.1.b) and zigzag (Fig.1.c) directions. In these models, both increasing distances between rings and two possible ring orientations (up and down) are considered. In particular, concerning the pattern along the armchair direction, (BN)$_3$ rings bonded together (model 2) as depicted in Fig.1.b, then separated by one benzene ring (model 3), and lastly by two benzene rings (model 4) are considered. Similarly, along the zigzag direction, two different distances between (BN)$_3$ rings are considered: borazine rings separated by a single C-chain (model 5) as reported in Fig.1.c and by three C-chains (model 6). Moreover, in models containing two borazine rings, both the parallel (up$\uparrow$-up$\uparrow$, see Fig.1.b,c) and anti-parallel (up$\uparrow$-down$\downarrow$, not shown here) orientations are investigated. All these atomistic models are illustrated in Supplementary Information. After structural optimizations, planarity is found to be preserved in each BN-ring doped model. In order to evaluate the relative stability of borazine-doped graphene, the cohesive energy of the six models is calculated using Eq.1, and presented in Fig.2 with a direct comparison with benchmark ideal BNC structures.

\begin{figure*}[!h]
\includegraphics[width = 0.5\textwidth]{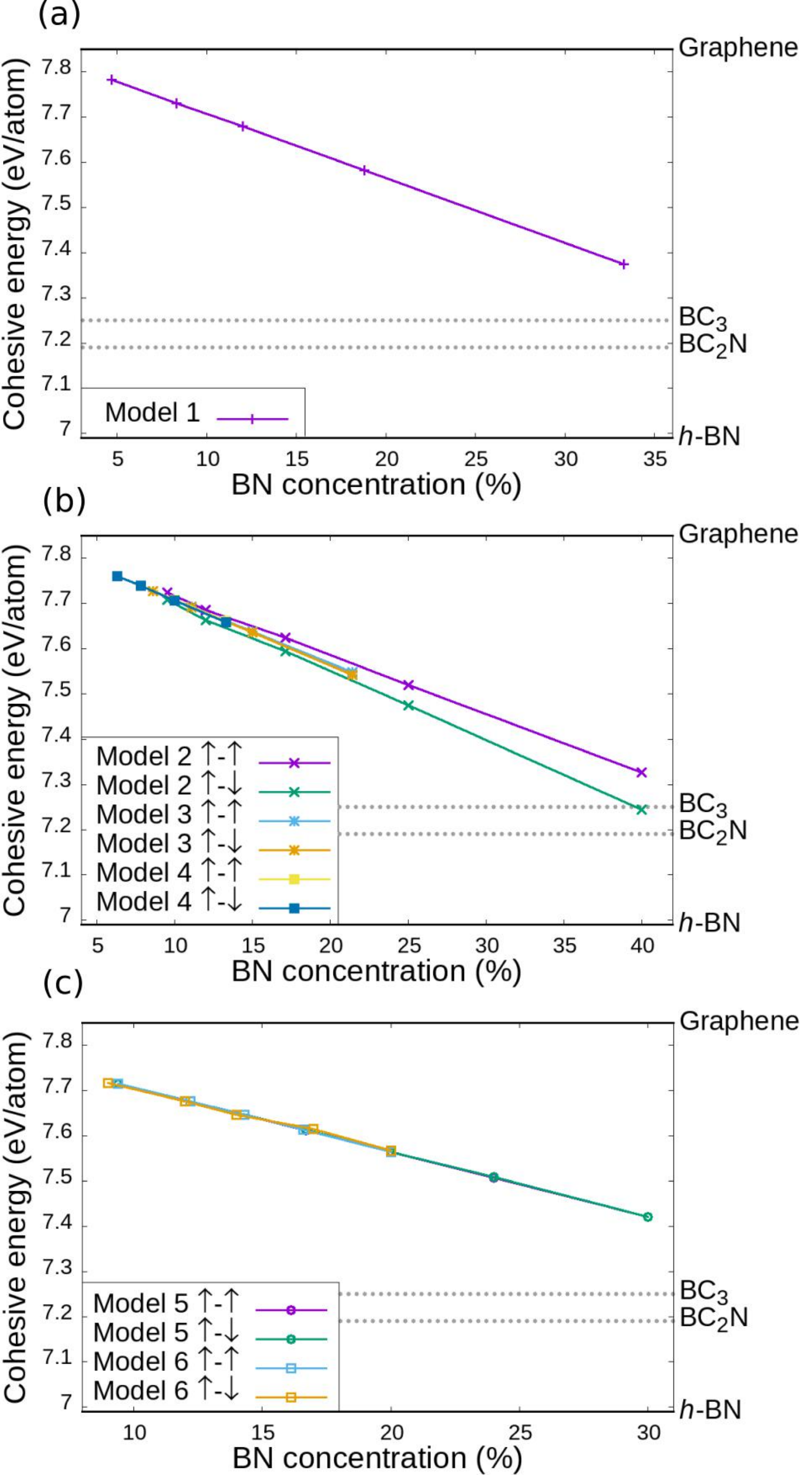}
\caption{Figure 2: Cohesive energies of borazine-doped graphene in function of BN-ring concentration for (a)  single borazine doping (model 1), (b) doping pattern formed by two (BN)$_3$ along the amrchair direction (models 2-4), (c) doping pattern formed by two (BN)$_3$ along the zigzag direction (models 5 \& 6). The reference energies are the cohesive energies of $\textit{h}$-BN as minimum and pristine graphene as maximum. BC$_3$ and BC$_2$N cohesive energies are indicated with horizontal dashed lines.}
\label{fig_sim2}
\end{figure*}
As expected, all the calculated cohesive energies of borazine-doped graphene are lower than the reference energy of pristine graphene, but almost everytime higher than BC$_3$ when considering large BN-ring concentration and always higher than C$_3$N$_4$ and BC$_2$N ideal systems. In particular, for each model regardless of its doping parameters, the cohesive energy value always lies in between the values of graphene and $\textit{h}$-BN. Consequently, each of the considered atomistic models represents a metastable structure that could potentially be synthesized and optimized to be implemented in devices. In particular, the cohesive energy trend line approaches the one of pristine graphene with a linear behavior with respect to the BN concentration with the system becoming less and less stable when the concentration of BN-ring increases (see Fig.2.a). In model 2, where the two rings are bonded together, a larger difference is observed. Indeed, in the up-up orientation, the two borazine rings are bonded through a N-B bond while in the up-down orientation, with a N-N bond. The difference in energy can be easily addressed to the lower binding energy of the N-N bond compared to the B-N one. This difference in cohesive energy is found to be larger at high BN concentrations but converges towards the same values at lower concentrations. Nevertheless, the largest energy difference in model 2 with different orientations is 0.08eV/atom (see Fig.2.b) suggesting that the cohesive energy is qualitatively insensitive to patterns and orientations considered herewith, as also confirmed in Fig.2.c. Moreover, the trend of the cohesive energy for different families of BNC monolayers has been evaluated in terms of the average distance between BN doping patterns; where an asymptotic behavior is observed towards the cohesive energy of pristine graphene (as shown in Supplementary Information).

\begin{figure*}[!h]
\includegraphics[width = 0.55\textwidth]{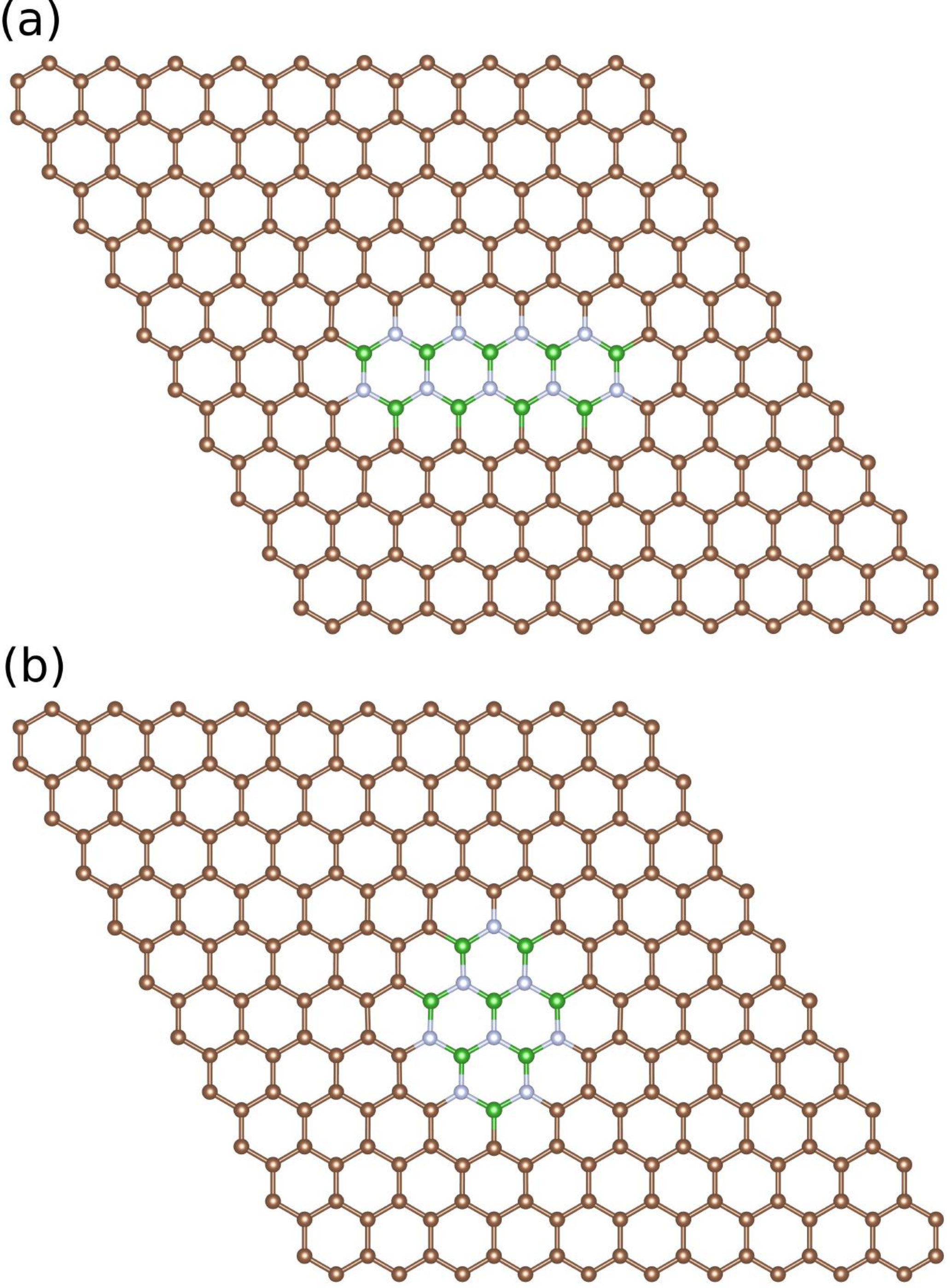}
\caption{Figure 3: Atomistic models of graphene doped with borazine (BN)$_3$ island patterns formed by four (BN)$_3$ rings (a) in a linear arrangment and (b) in a square configuration.}
\label{fig_sim3}
\end{figure*}

Lastly, the stability of the previous models has been compared to the one of isolated BN islands. In particular, BN islands with different shapes and concentrations have been studied. Considering BN pattern formed by four fused borazine rings firstly arranged in a linear shape (model 7 - Fig.3.a) and then condensed in a squared shape (model 8  - Fig.3.b), lowering the BN-C interface, the cohesive energy is decreased by just 0.01eV. Confronting the cohesive energy of models 7-8 with the former ones, one should naively expect fused ring patterns to be more stable due to the segregation trend as previously mentioned. However, the cohesive energy of BN island patterns is found to be comparable to those of isolated rings, i.e. for similar concentrations, the difference is $\sim$0.02eV/atom.

After considering their respective stabilities, the electronic properties of various BNC atomic structures have been calculated using both DFT approaches and beyond. $\textit{Ab initio}$ density of electronic states (DOS) of various graphene doped BN rings (models 1,2 and 5) with concentrations ranging from 12$\%$ -- 40$\%$ have been evaluated (Fig. 4) and band gaps values have been preliminarily estimated at the DFT-PBE level of theory.

\begin{figure*}[!h]
\includegraphics[width = 0.95\textwidth]{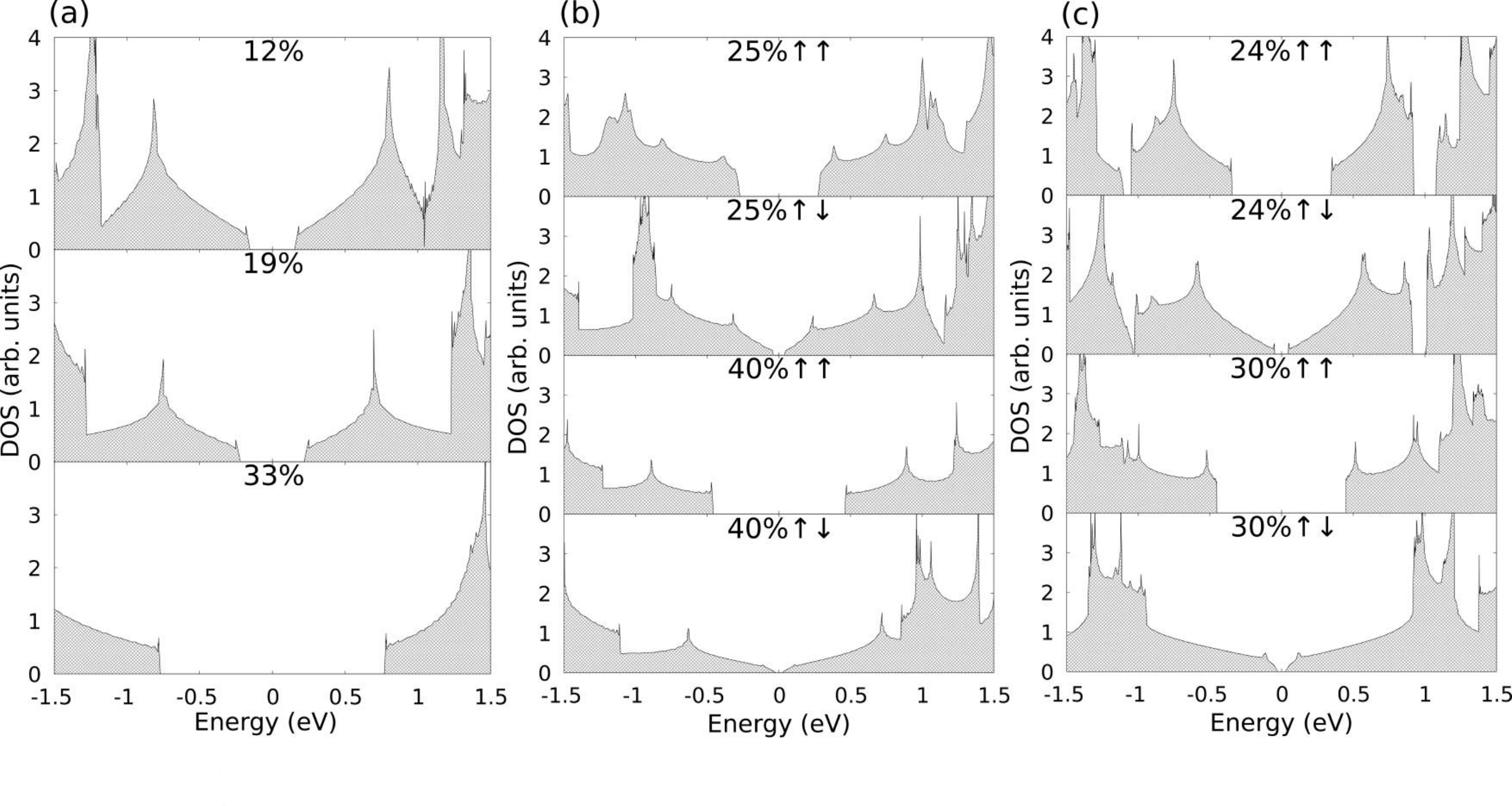}
\caption{Figure 4: Density of electronic states of (a) model 1, (b) model 2, and (c) model 3 with various concentrations as indicated in each frame. The parallel and anti-parallel orientations are also mentioned in (b) and (c). Zero energy is always located at the middle of the gap.}
\label{fig_sim4}
\end{figure*}

For each BN-doping structural arrangement and concentration, a band gap opening is observed when compared to pristine graphene. All these band gaps are found to be direct, similarly to other BN-doped graphene models considered in previous studies \cite{Shinde}. Direct band gaps present an advantage with respect to the indirect ones (as in several inorganic materials) since they could open up the possibility of exhibiting high absorption coefficient, as well as other useful optoelectronic properties. 
Regarding the composition of the valence and conduction bands related to these models, for every concentration considered, 2$p_z$ carbon orbitals contribution is prevalent in the DOS leading to the dominant graphene contribution to the electronic conduction. Due to the presence of the same number of Boron and Nitrogen atoms, there are no defect states in the electronic structures suggesting strong coupling between the carbon and doping atoms and the absence of any  spin-polarized phenomena.
Moreover, confronting the same models with the same concentration but with two different mutual ring orientations as illustrated in Fig.4b-c, parallel orientation models ($\uparrow$-$\uparrow$ case) are found to have a considerably higher band gap compared to the anti-parallel orientation ($\uparrow$-$\downarrow$ case) in the similar pattern. This can be linked to the chiral symmetry breaking previously discussed. In particular, two BN rings are considered in each unit cell in the models reported in Fig. 4b-c. The parallel orientation case will thus result in one of the sublattices containing only Carbon and Nitrogen atoms and the other sublattice containing only Carbon and Boron atoms, leading to an asymmetric state hence a large band gap opening. On the other hand, in the anti-parallel case, both sublattices are composed by the same number of Carbon, Nitrogen and Boron atoms resulting in a symmetric state hence a small band gap, as observed in Fig. 4. This implies that, during the design process of BNC nanomaterials, particular attention needs to be put on the orientation parameter because of its strong influence on the electronic properties and the possibility of a band gap closing.

However, DFT is well known to be a ground state theory and 
that electronic band structures, in particular the size of band gaps, are not well reproduced by most exchange-correlation functionals \cite{dftproblem,sigw}.
Consequently, the band gap values calculated herewith employing this technique are strongly underestimated. Even if a qualitative comparison of DFT band gap values is still possible between BNC models featuring more or less the same number and type of atoms, the $ab$ $initio$ values are essentially wrong and can not be directly compared to experimental band gaps. In order to correct these values, beyond-DFT techniques are employed using many-body perturbation theory (MBPT). More specifically, some parallel oriented models have been selected to calculate the quasiparticle (G$_0$W$_0$) corrections to their band gap values (Fig. 5) \cite{Yambo1}.

\begin{figure*}[!h]
\includegraphics[width = 0.6\textwidth]{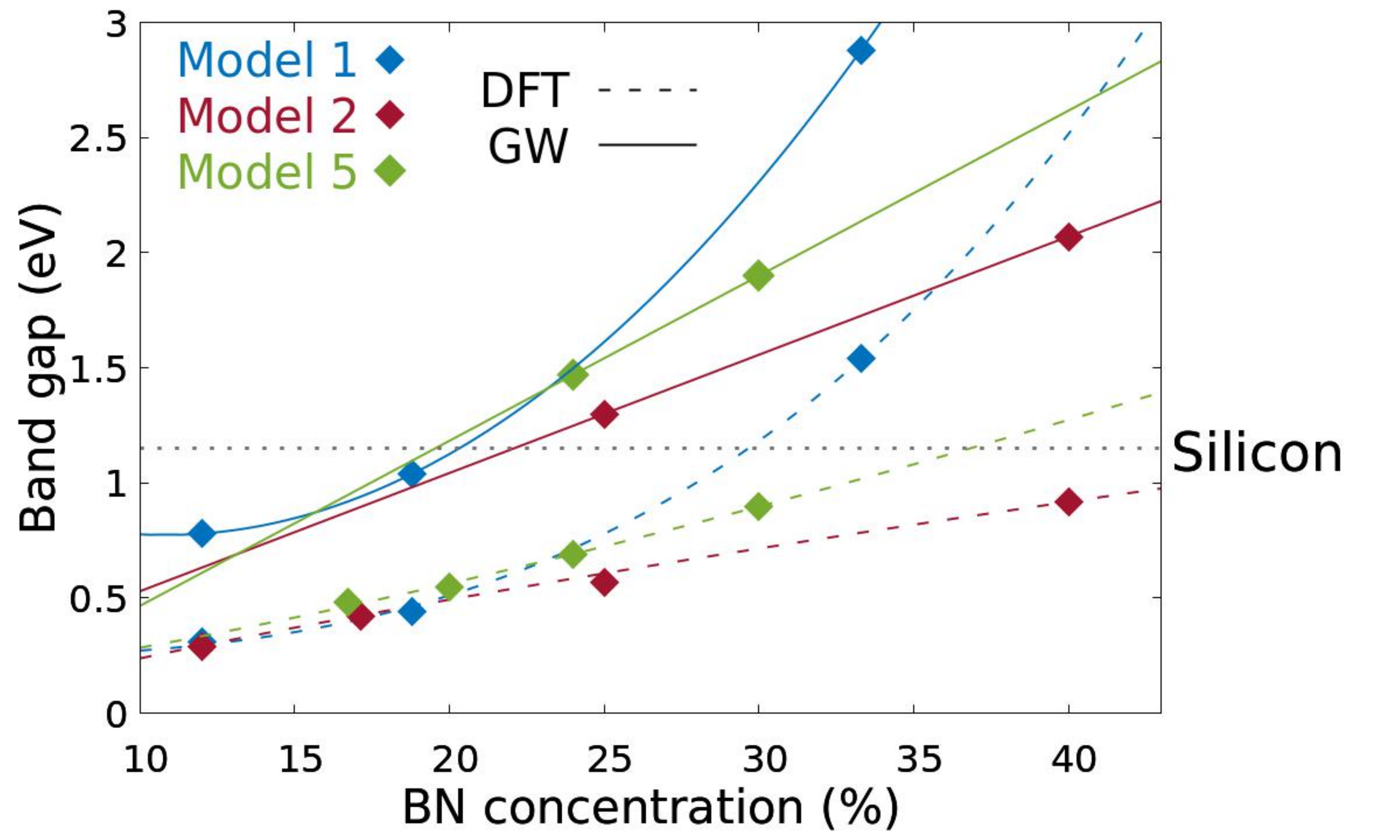}
\caption{Figure 5: Band gap values calculated within DFT-PBE (dashed lines) and within the GW approximation (solid lines) for models 1, 2 and 5 represented in blue, red and green, respectively. Curves fitting band gap values are represented as a guide for the eyes, highlighting the band gap enhancement when BN concentration is increased. The GW band gap of Silicon is indicated as a reference using a horizontal dashed line.}
\label{fig_sim5}
\end{figure*}

As expected, for each BNC model, the GW bandgap values are found to be significantly larger than DFT predictions, with corrections in the range of 0.5 - 1.3 eV. In contrast to the cohesive energy study, the band gap values strongly depend not only on the orientation, as already discussed, but also on the BN-doping pattern. In fact, since the band gap evolution trend is different for each model considered herewith, high BN-concentration does not necessarily mean large band gap (see Fig. 5). Differently from other methodologies used to modify graphene electronic properties \cite{substrate}, a quite wide range of band gap values is obtained highlighting the possibility of fine-tuning graphene by proper control of doping pattern and parameters. In particular, when considering parallel orientation between rings, band gaps close to the one of bulk Silicon (i.e. 1.15 eV calculated with the GW approximation) can be achieved. Indeed, a band gap close to Silicon bulk is obtained when considering the model featuring a single borazing ring (i.e. model 1) with BN-concentration of 19$\%$ or featuring bonded borazines with a B-N bond (i.e. model 2 with parallel ring orientation) with BN-concentration of 25$\%$. When considering non-bonded rings in similar concentration (model 5), even if the band gap is slightly raised, it still presents only a 0.32 eV band gap difference with bulk Silicon. Moreover, band gaps in the visible range can be achieved when considering high concentration models, opening up the possibility to use hybrid BNC materials as wide-band-gap semiconductors in opto-electronic devices. On the other hand, smaller band gaps could be required for applications in infrared regimes. These smaller gaps are found in antiparallel-ring doping patterns or in small concentrations models. Unfortunately, even if MBPT techniques produce accurate band gap values that are comparable to photo-emission and photo-absorption experiments, the high computational cost required to calculate the microscopic dielectric function to apply the quasi-particle correction to the DFT band gap values limits the application of GW methods to BNC models featuring small band gaps hence containing a large number of atoms. Consequently, one could search for a compromise between high accuracy band gap values, as provided by the GW technique, and low computational cost, as provided by DFT techniques. This could be achieved by using hybrid functionals, i.e. including part of the exact exchange from Hartree-Fock method in the exchange-correlation functionals used in DFT methods. In particular, global hybrids add a reasonable fraction of the full nonlocality to the calculation within the exchange \cite{gh}. In these functionals, the most computational challenging part is the calculation of the slowly decaying exchange with the distance. Therefore, short-range screened hybrids are nowadays among the most used hybrid functional used  \cite{HSE3} \cite{HSErev} due to the inclusion of Hartree-Fock exchange only in the short-range part of the electron-electron interaction, instead of evaluating the non-local Hartree-Fock part of the exchange. However, numerous hybrid functionals are available depending on the parametrization of the functional and their accuracy could strongly vary depending on the system under study. Among the short-range hybrid functions, Heyd-Scuseria-Ernzerhof (HSE) hybrids \cite{HSEb} have been extensively applied for the calculation of band gap of semiconductors demonstrating a higher accuracy \cite{HSEaccuracy} compared to other types of hybrid functional as well as moderate computational cost. Here, different techniques have been employed to evaluate the band gap values of BNC hybrid monolayers to assess the accuracy of hybrid functions in order to search for a reasonable alternative to the GW approximation in models featuring a large number of atoms. More specifically, both short-range screened hybrid HSE06 and global hybrid PBE0 \cite{PBE0} band gap values have been calculated and directly compared with the high-accuracy GW values (see Table 1).

\begin{table}
\captionof{table}{Table I: Band gap values (in eV) calculated with DFT-PBE, HSE06, PBE0 and GW approaches for models 1, 2 and 5 in the parallel orientation and with different concentrations.}
\begin{tabular}{|c|ccccc|}
\hline
\multicolumn{1}{|l|}{}   & \multicolumn{1}{c|}{\begin{tabular}[c]{@{}c@{}}BN  (\%)\end{tabular}} & \multicolumn{1}{l|}{DFT - PBE (eV)} & \multicolumn{1}{l|}{HSE06 (eV)} & \multicolumn{1}{l|}{PBE0 (eV)} & \multicolumn{1}{l|}{GW (eV)} \\ \hline
\multirow{3}{*}{Model 1} & 12                                   & 0.31                                & 0.45                            & 0.50                           & 0.77                         \\ \cline{2-6}
                         & 19                                   & 0.44                                & 0.64                            & 0.84                           & 1.04                         \\ \cline{2-6} 
                         & 33                                   & 1.54                                & 2.15                            & 2.69                           & 2.88                         \\ \hline 
\multirow{2}{*}{Model 2} & 25                                   & 0.57                                & 0.86                            & 1.18                           & 1.30                         \\ 
                         & 40                                   & 0.92                                & 1.35                            & 1.80                           & 2.07                         \\ \hline
\multirow{2}{*}{Model 5} & 24                                   & 0.69                                & 0.99                            & 1.33                           & 1.47                         \\ 
                         & 30                                   & 0.90                                & 1.30                            & 1.74                           & 1.90                         \\ \hline
\end{tabular}

\end{table}
\medskip
As explained above, DFT values are largely underestimating the GW band gaps while both hybrid functional values always lie in between the two predictions. Despite there is no consistent discrepancy between accurate GW band gaps and the two hybrid functionals considered herewith, PBE0 band gap values are found to be generally closer to the GW ones (see Table 1), with differences in the range of 0.1-0.3 eV. On the other hand, HSE06 differences with GW values are found to be larger when the BN ring concentration is raised, in fact the maximum difference (0.7 eV) is found at the highest concentrations (40$\%$ and 33$\%$) while the lowest difference (0.3 eV) is found at the lowest concentration (12$\%$). Thanks to the strong band gap enhancement at large concentrations and the relative close values obtained using PBE0, hybrid functionals represent a viable alternative to GW approximation for the calculation of band gap values. By lowering the BN concentration, hybrid functionals still provide enhanced band gaps compared to DFT allowing less expensive calculations on very large supercells. However, GW approximation still provides strongly enhanced band gap values compared to DFT and, at the lowest concentrations, to hybrid functionals, thus remaining the leading methodology for band gap calculations.\\
Until now, periodically doped BNC monolayers have been considered since they are the targeted materials in bottom-up synthetic approaches. However, perfect control of the system periodicity is still practically questionable. Therefore, even though they could be weak, disorder effects on the electronic properties are computed by tight-binding calculations (Fig.6) for the disordered systems built starting from the model containing a single borazine ring (see Fig.1a). Here, two types of disorder (i.e., location and rotation) are considered. In particular, in the first case, a number of BN rings are displaced from their original position in the periodic system to some neighbouring positions while, for the second case, a number of BN rings are randomly rotated to be anti-parallel to their original orientation. In addition, different disorder probabilities (i.e., percentages of BN-rings displaced or rotated) were considered. The detailed description of the tight-binding calculations and disorder models are presented in the Supplementary Information.
\begin{figure*}[!h]
\includegraphics[width = 0.8\textwidth]{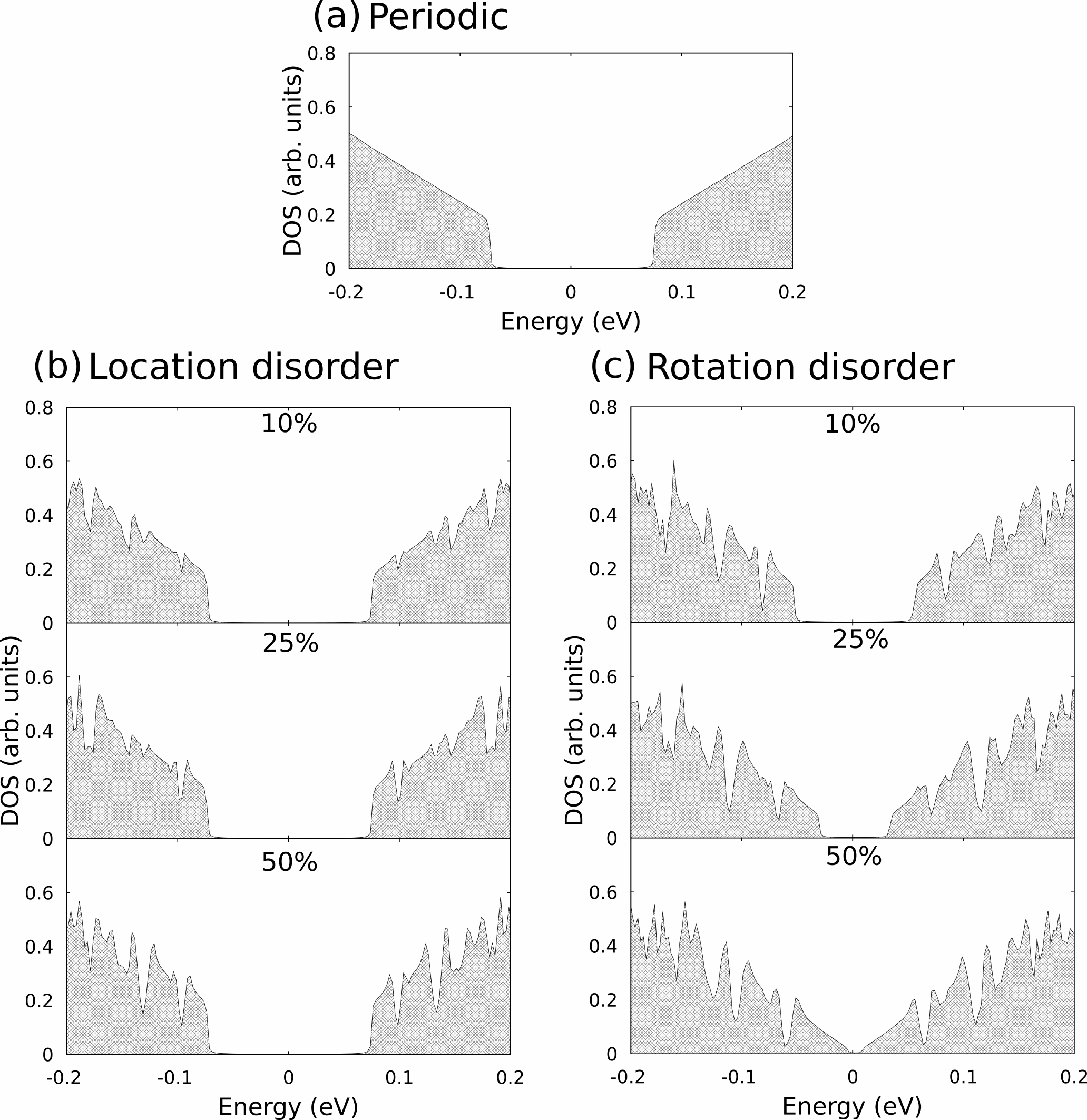}
\caption{Figure 6: Electronic density of states of (a) periodic, (b) location and (c) rotation disordered models computed by tight-binding calculations. Each model has been built using the original 8$\times$8 cell containing a single borazine ring (see Fig.1a). Moreover, different disorder concentrations have been considered as indicated in each frame. Location disorder models have been built using the model M$_2$ described in the Supplementary Information. In (c), the rotation disorder is added in the location disordered system in (b) with 25$\%$ of displaced BN rings.}
\label{fig_sim5}
\end{figure*}

Interestingly, the tight-binding simulations show that the location disorder (see also results with different displacement distances additionally presented in Fig.S9 in the SI) likely does not present any significant effect on the formation of bandgap. This could be essentially explained by the feature that the graphene sublattice asymmetry (i.e., symmetry breaking) is the key ingredient for bandgap opening. In particular, as explained by DFT studies above, in the periodic case B-atoms (N-atoms) replace C-atoms in one graphene sublattice (other sublattice, respectively), leading to a certain level of sublattice asymmetry and hence a finite bandgap. When the location disorder is applied, the sublattice asymmetry level is unchanged, excepting that distance between BN-rings spatially varies. In these cases (i.e., both periodic and aperiodic systems), the sublattice asymmetry level is tuned when changing the BN-ring concentration, as presented above, i.e., the dependence of bandgap on the BN-ring concentration is obtained. The situation however drastically changes when the rotation disorder occurs, i.e., the bandgap is significantly reduced when increasing this disorder. Note that increasing the rotation disorder while the BN-ring concentration is fixed, the system transits from the strongest sublattice asymmetry state (0$\%$ rotation) to the sublattice symmetry one (50$\%$ rotation). Thus, the sublattice asymmetry level is reduced when increasing this disorder, which is obviously the essential reason of the bandgap reduction observed. The 50$\%$ rotation disorder here is actually similar to the situations when a pair of up-down rings in the unit cell is invesitated above by the DFT calculations, i.e., a neglegible (or small) bandgap is indeed obtained. Thus, tight-binding calculations confirm all main features predicted by DFT and clearly illustrate that graphene sublattice symmetry breaking is the essential ingredient for bandgap opening. The obtained results suggest that controlling BN-ring concentration as well as ring rotation is the key to tune the bandgap whereas perfect periodicity of the location of the dopant is not strictly required.

\section{Conclusion}
The present work reports on BN-doping effect in both structural and electronic properties of 2D graphene. In all cases, BNC materials exhibit similar stability when compared to hypothetical 2D benchmark systems, highlighting the possibility to synthesize a specifically tuned hybrid BNC monolayer using for example bottom-up chemical approaches as for the on-surface synthesis of atomically precise graphene nanoribbons \cite{cai}\cite{cai2}\cite{ruffieux}. However, the band gap value is shown not only to vary significantly with the BN ring concentration but also to be dependent on ring orientations. Indeed, due to the strong chiral symmetry breaking, the obtained band gap is larger in the parallel oriented ring cases than that in the anti-parallel ones with the same doping concentration. Moreover, GW approximation has been employed to calculate corrections to the DFT band gap of selected BNC monolayers exhibiting band gaps close to the one of Silicon. The optimal BN-ring concentration in graphene is found to be close to 20-25$\%$. In addition, BNC models have the advantage of presenting direct band gaps which is quite interesting for optoelectronics. 
Finally, when comparing two of the most frequently used hybrid functionals with the GW results, PBE0 is found to predict more accurate results in comparison with HSE06 for every investigated BNC model. Furthermore, thanks to the band gap enhancement particularly high at larger concentrations, hybrid functionals open up the possibility to use less demanding calculations for the evaluation of band gap values for large supercell BNC models. Lastly, in the disordered systems, the bandgap is shown to negligibly depend on the position of BN rings but it is significantly affected by the rotation disorder.
In summary, relatively stable 2D materials could possibly be achieved by fusing graphene with BN-domains featuring direct effective band gap opening that can assume a wide range of values opening up the possibility of multiple optoelectronic applications. 

\medskip
	
	\textbf{Acknowledegments} - The authors acknowledge fundings from the European Union's Horizon 2020 Research and Innovation programme under the Marie Sklodowska-Curie entitled STiBNite (N$^{\circ}$ 956923), from the European Union’s Horizon 2020 Research Project and Innovation Program — Graphene Flagship Core3 (N$^{\circ}$ 881603), from the F\'ed\'eration Wallonie-Bruxelles through the ARC on Dynamically Reconfigurable Moir\'e Materials (N$^{\circ}$ 21/26-116), from the Flag-Era JTC projects “TATTOOS” (N$^{\circ}$ R.8010.19) and “MINERVA” (N$^{\circ}$ R.8006.21), from the EOS project “CONNECT” (N$^{\circ}$ 40007563) and from the Belgium F.R.S.-FNRS through the research project (N$^{\circ}$ T.029.22F). Computational resources have been provided by the CISM supercomputing facilities of UCLouvain and the C\'ECI consortium funded by F.R.S.-FNRS of Belgium (N$^{\circ}$ 2.5020.11).

\newpage

\clearpage

\title{Supplemental Material for:\\
First-principles study of the structural and electronic properties BN-ring doped graphene}
\author{Laura Caputo, Viet-Hung Nguyen and Jean-Christophe Charlier}
\affiliation{Institute of Condensed Matter and Nanosciences, Universit\'{e} catholique de Louvain (UCLouvain), B-1348 Louvain-la-Neuve, Belgium}

{
\let\clearpage\relax
\maketitle
}

\section{Atomistic models}
In figure 1 (main text), model 2 and model 5 are depicted to represent the BN-doping pattern along the armchair and zigzag directions, respectively, both in a parallel orientation. In Figs.S1-S3, models 3,4 and 6 are reported in both parallel and antiparallel orientations.\\

\begin{center}
\centering
\includegraphics[width = 0.65\textwidth]{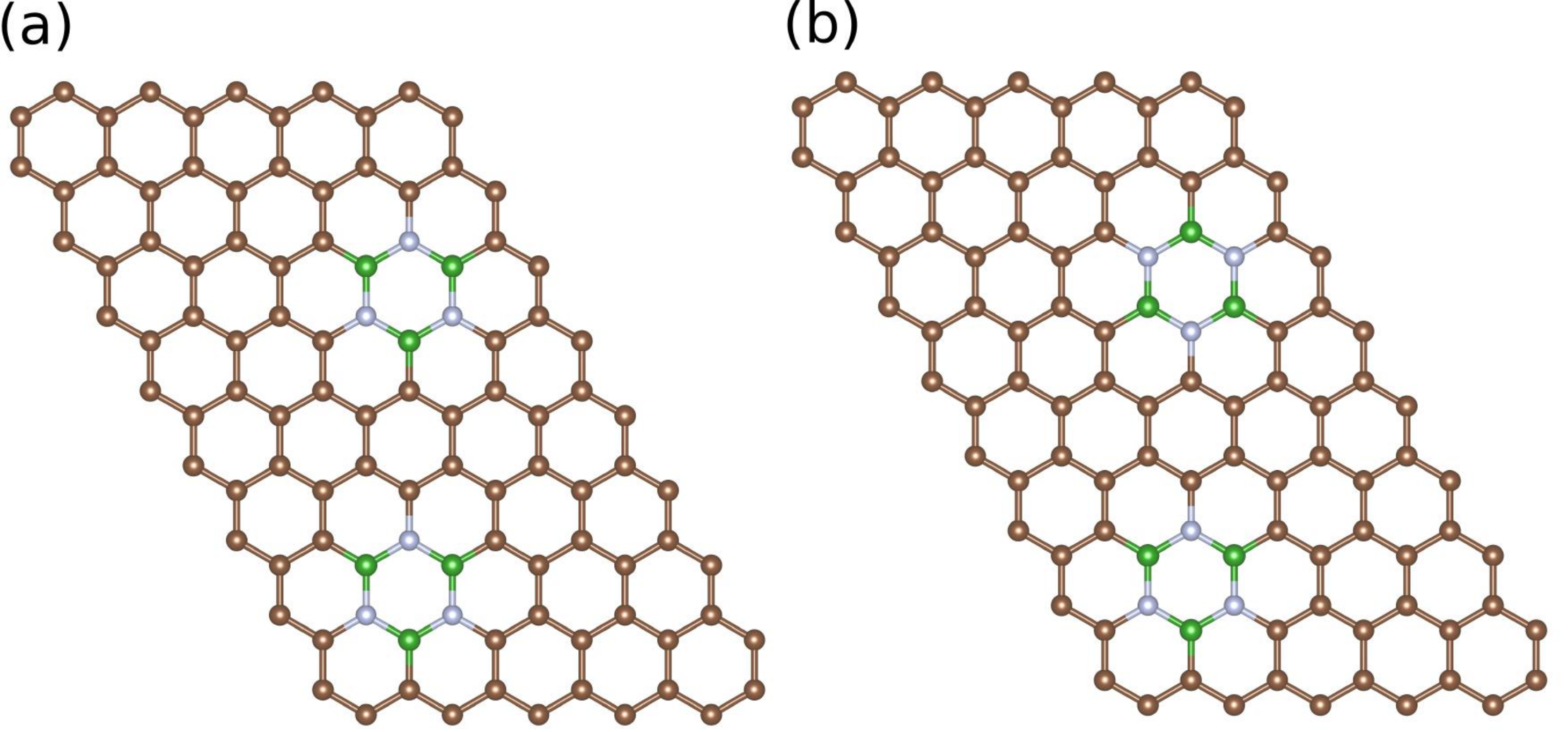}
\captionof{figure}{Figure S1: Atomistic model 3 in (a) up-up and (b) up-down orientations.}
\end{center}

\begin{center}
\centering
\includegraphics[width = 0.7\textwidth]{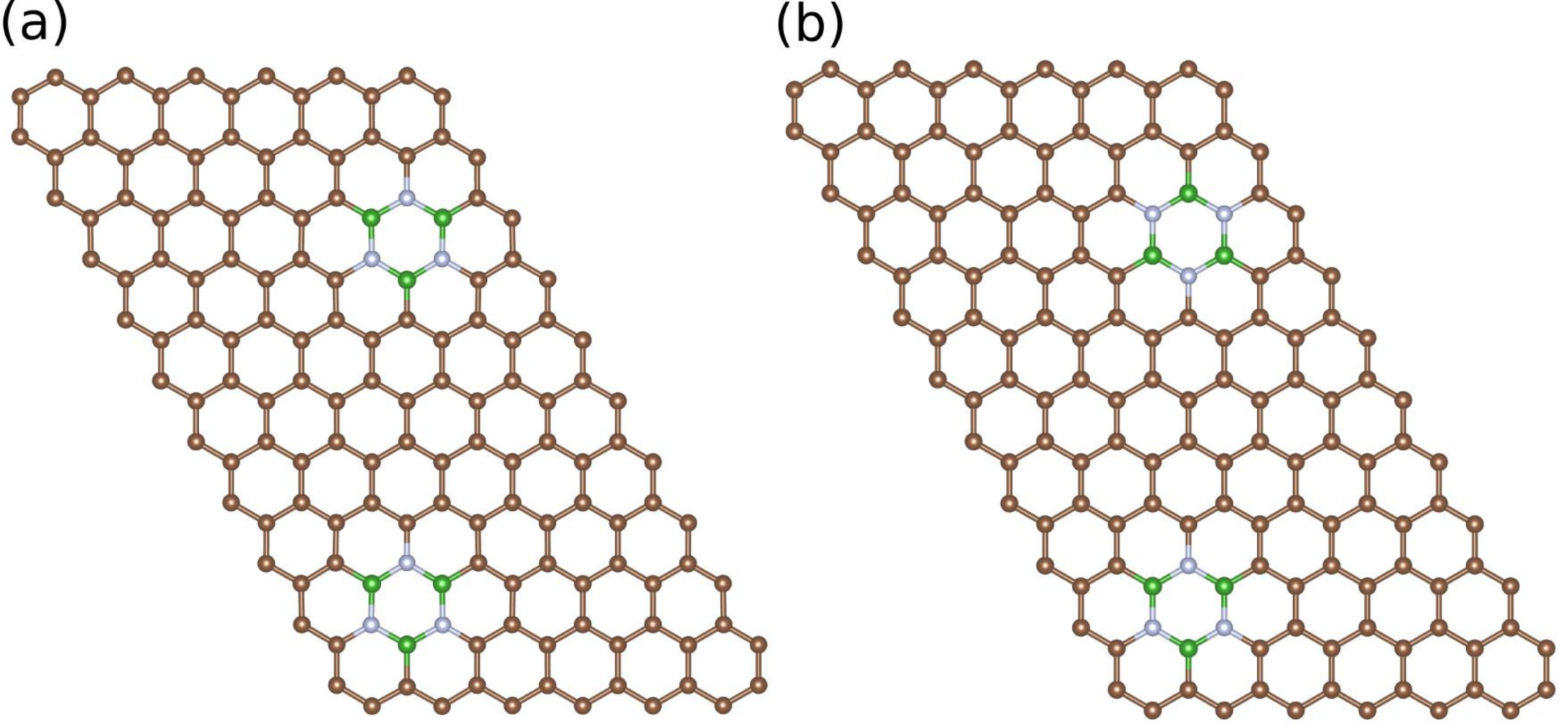}
\captionof{figure}{Figure S2: Atomistic model 4 in (a) up-up and (b) up-down orientations.}
\end{center}

\begin{center}
\centering
\includegraphics[width = 0.7\textwidth]{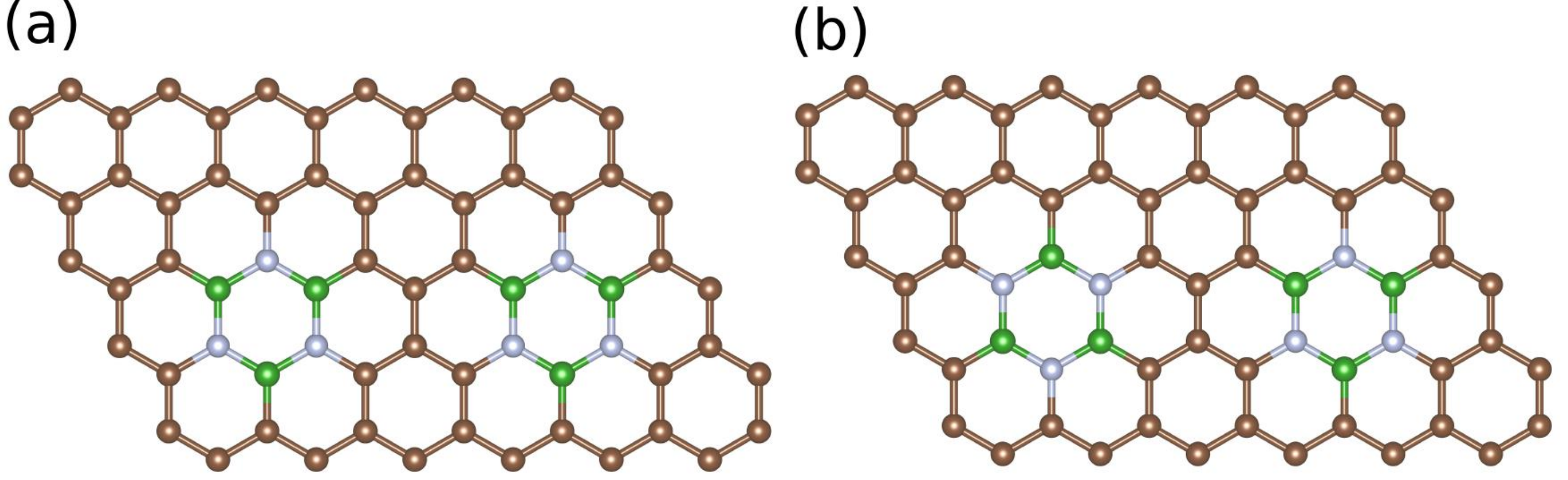}
\captionof{figure}{Figure S3: Atomistic model 6 in (a) up-up and (b) up-down orientations.}
\end{center}

In Fig.2 (main text), the stability of BNC materials is compared to different benchmark materials. The atomistic models of these are reported in Fig. S4.

\begin{center}
\centering
\includegraphics[width = 0.8\textwidth]{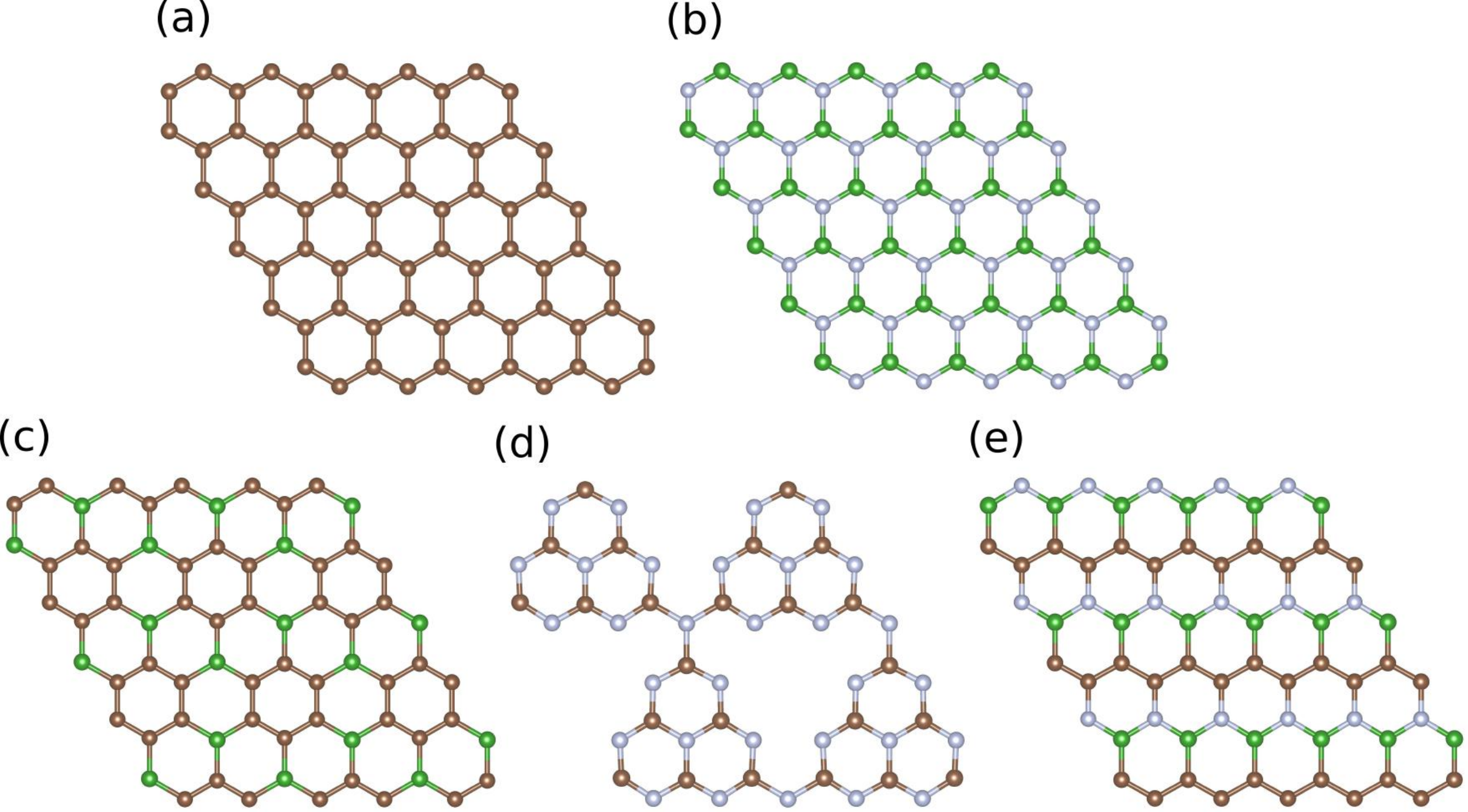}
\captionof{figure}{Figure S4: Atomistic models for (a) graphene, (b) \textit{h}-BN, (c) BC$_3$, (d) C$_3$N$_4$, (e) BC$_2$N.}
\end{center}

\section{Cohesive energy of BNC monolayers}
Cohesive energies for each model considered in both orientations are reported in Table SI. 

\begin{table}[]
\begin{tabular}{|cc|ccc|ccc|}
\hline
\multicolumn{2}{|c|}{Model 1}          & \multicolumn{3}{c|}{Model 2}                                        & \multicolumn{3}{c|}{Model 3}                                        \\ \hline
\multicolumn{1}{|c|}{BN (\%)} & E (eV) & \multicolumn{1}{c|}{BN (\%)} & \multicolumn{1}{c|}{E$_{\uparrow \uparrow}$ (eV)} & E$_{\uparrow \downarrow}$ (eV) & \multicolumn{1}{c|}{BN (\%)} & \multicolumn{1}{c|}{E$_{\uparrow \uparrow}$ (eV)} & E$_{\uparrow \downarrow}$ (eV) \\ \hline
\multicolumn{1}{|c|}{5}       & 7.78   & \multicolumn{1}{c|}{10}      & \multicolumn{1}{c|}{7.72}   & 7.71   & \multicolumn{1}{c|}{9}       & \multicolumn{1}{c|}{7.73}   & 7.73   \\ \hline
\multicolumn{1}{|c|}{8}       & 7.73   & \multicolumn{1}{c|}{12}      & \multicolumn{1}{c|}{7.69}   & 7.66   & \multicolumn{1}{c|}{11}      & \multicolumn{1}{c|}{7.69}   & 7.69   \\ \hline
\multicolumn{1}{|c|}{12}      & 7.68   & \multicolumn{1}{c|}{17}      & \multicolumn{1}{c|}{7.62}   & 7.59   & \multicolumn{1}{c|}{15}      & \multicolumn{1}{c|}{7.64}   & 7.64   \\ \hline
\multicolumn{1}{|c|}{19}      & 7.58   & \multicolumn{1}{c|}{25}      & \multicolumn{1}{c|}{7.52}   & 7.48   & \multicolumn{1}{c|}{21}    & \multicolumn{1}{c|}{7.55}   & 7.54   \\ \hline
\multicolumn{1}{|c|}{33}      & 7.37   & \multicolumn{1}{c|}{40}      & \multicolumn{1}{c|}{7.33}   & 7.24   & \multicolumn{3}{c|}{}                                               \\ \hline
\end{tabular}
\medskip
\\
\begin{tabular}{|ccc|ccc|ccc|}
\hline
\multicolumn{3}{|c|}{Model 4}                                        & \multicolumn{3}{c|}{Model 5}                                        & \multicolumn{3}{c|}{Model 6}                                        \\ \hline
\multicolumn{1}{|c|}{BN (\%)} & \multicolumn{1}{c|}{E$_{\uparrow \uparrow}$ (eV)} & E$_{\uparrow \downarrow}$ (eV) & \multicolumn{1}{c|}{BN (\%)} & \multicolumn{1}{c|}{E$_{\uparrow \uparrow}$ (eV)} & E$_{\uparrow \downarrow}$ (eV) & \multicolumn{1}{c|}{BN (\%)} & \multicolumn{1}{c|}{E$_{\uparrow \uparrow}$ (eV)} & E$_{\uparrow \downarrow}$ (eV) \\ \hline
\multicolumn{1}{|c|}{6}       & \multicolumn{1}{c|}{7.76}   & 7.76   & \multicolumn{1}{c|}{9}       & \multicolumn{1}{c|}{7.72}   & 7.72   & \multicolumn{1}{c|}{9}       & \multicolumn{1}{c|}{7.72}   & 7.72   \\ \hline
\multicolumn{1}{|c|}{8}       & \multicolumn{1}{c|}{7.74}   & 7.74   & \multicolumn{1}{c|}{17}      & \multicolumn{1}{c|}{7.61}   & 7.61   & \multicolumn{1}{c|}{12}      & \multicolumn{1}{c|}{7.68}   & 7.68   \\ \hline
\multicolumn{1}{|c|}{10}      & \multicolumn{1}{c|}{7.71}   & 7.71   & \multicolumn{1}{c|}{20}      & \multicolumn{1}{c|}{7.56}   & 7.57   & \multicolumn{1}{c|}{14}      & \multicolumn{1}{c|}{7.65}   & 7.65   \\ \hline
\multicolumn{1}{|c|}{13}      & \multicolumn{1}{c|}{7.66}   & 7.66   & \multicolumn{1}{c|}{24}      & \multicolumn{1}{c|}{7.51}   & 7.51   & \multicolumn{1}{c|}{17}      & \multicolumn{1}{c|}{7.61}   & 7.62   \\ \hline
\multicolumn{1}{|c|}{}        & \multicolumn{1}{c|}{}       &        & \multicolumn{1}{c|}{30}      & \multicolumn{1}{c|}{7.42}   & 7.42   & \multicolumn{1}{c|}{20}      & \multicolumn{1}{c|}{7.57}   & 7.57   \\ \hline
\end{tabular}
\captionof{table}{Table SI: Cohesive energies of borazine-doped graphene with various concentrations from Model 1 to Model 6 in both parallel and anti-parallel orientations.}
\newpage
\end{table}
The cohesive energies of BNC monolayer in function of the average distance between BN doping patterns are found to exhibit an asymptotic behaviour, approaching the graphene cohesive energy value as the distance between patterns is raised, i.e. the concentration is lowered (see Fig. S5). This behaviour is particularly well observed when the vertical alignment between rings is considered, while the trend is slightly oscillating when the horizontal alignment is considered, probably due to the size of the supercells used. In Fig. S5b, model 2 is the only one where the different orientation between rings gives different stability, due to the fact that the rings are bonded together. Model 3 and 4, on the contrary, do not show any difference considering different orientation between rings. In figure S5c, the cohesive energy in function of the average BN distance is found to be almost insensitive also to the distance between rings.
\begin{figure*}
\centering
\includegraphics[width = 0.56\textwidth]{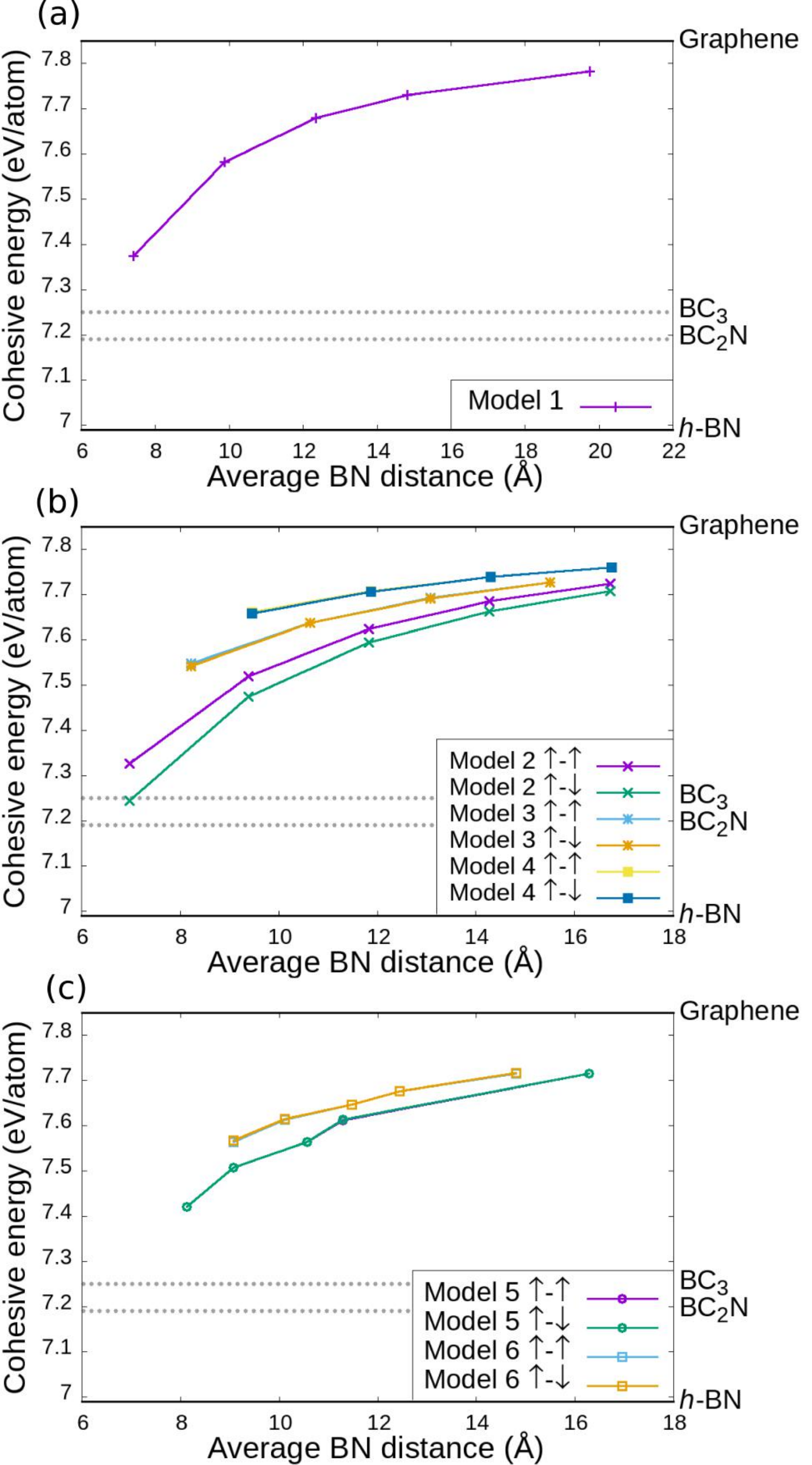}
\caption{Figure S5: Cohesive energies of borazine-doped graphene in function of the average distance between BN doping patterns for: (a) single borazine doping (model 1), (b) two borazine rings doping pattern with vertical alignment (models 2, 3 and 4), and (c) two borazine rings doping pattern with horizontal alignment (models 5 and 6). The reference energy is the cohesive energy of pristine graphene. BC$_3$ and BC$_2$N cohesive energies are indicated with horizontal dashed lines.}
\label{fig_sim5}
\end{figure*}

\section{Band gaps}

Comparison between GW and hybrid functional calculations for the band gap in BNC materials.

\begin{figure*}[!h]
\centering
\includegraphics[width = 0.8\textwidth]{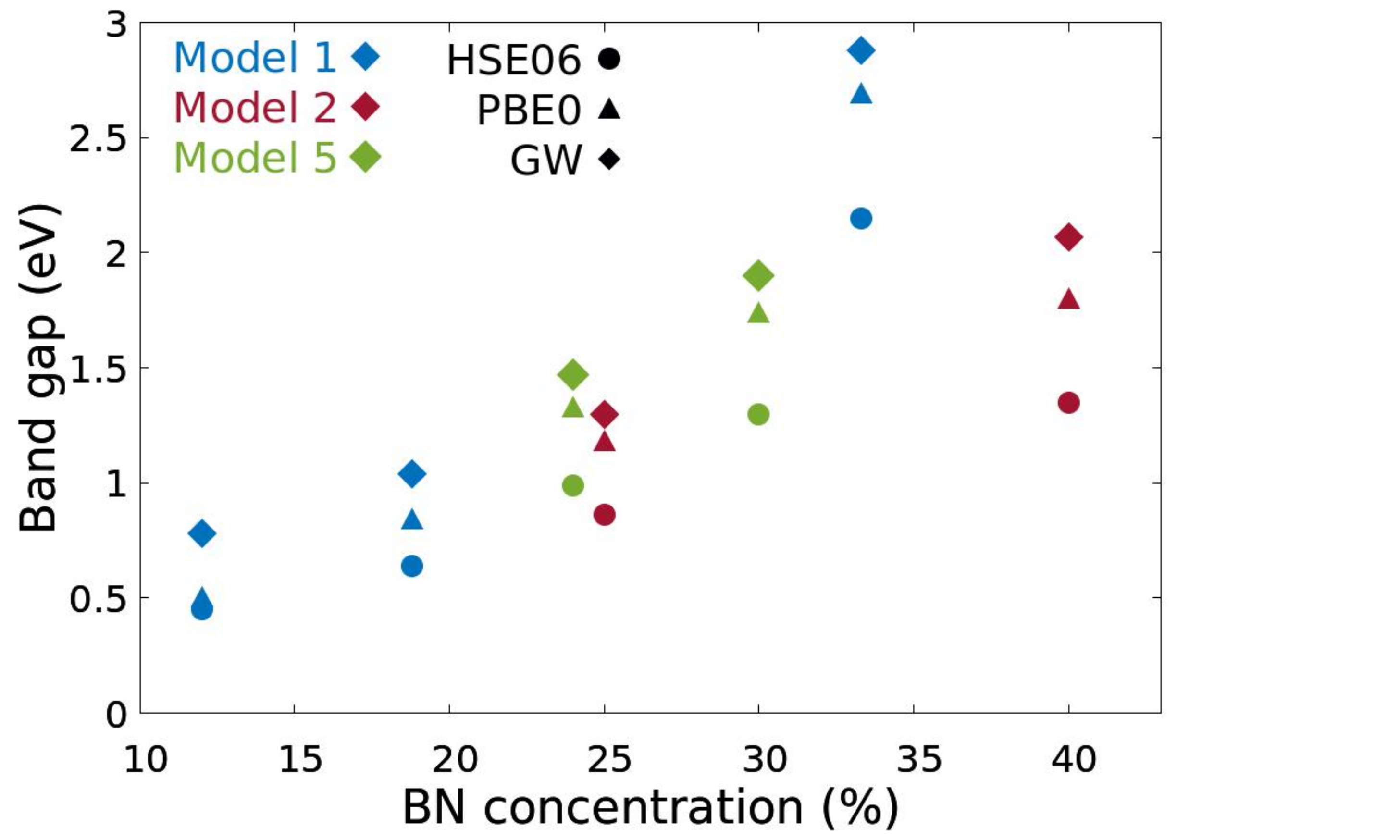}
\caption{Figure S6: Comparison between band gap values for model 1, 2 and 5 represented in blue, red and green, respectively. HSE06, PBE0 and GW estimations are represented by circles, triangles and rhombuses, respectively.}
\label{fig_sim6}
\end{figure*}

\section{Aperiodic effects}

Even though they are expected to be weak, the aperiodic (i.e., disorder) effects are computed and presented in this section. In particular, two typical disorders related to either ring location or ring orientation are considered in this work. The location disorder occurs when some BN-rings are not placed in the correct position satisfying the periodic condition (see Fig.S7). The latter disorder is related to ring rotation occurring when in a system of parallel-oriented rings, some rings are incorrectly replaced by anti-parallel ones. To compute the effects of these disorders, largely extended cells (i.e., large supercells) are first created by periodically repeating a large number of original cells investigated in the DFT calculations. Then, location disorder is introduced by allowing a number of BN rings to move from their original position to neighbouring positions as shown in Fig. S7 with displacements 1,2 and 3. Here, we present only calculations for extended supercells created by the original 8$\times$8 cell containing a single borazine ring (see Fig.1a in the main text). As mentioned in the main text, the main aim of the present work is to investigate periodic systems that could be produced by the bottom-up techniques, therefore we limited our investigation on only these small ring displacements. In addition, three different models are studied: Model M$_1$ featuring displacement 1, model M$_2$ featuring displacements 1 and 2, and lastly model M$_3$ featuring all displacements 1, 2 and 3. At last, in each case, different disorder probabilities (i.e., different numbers of BN rings displaced from their original position) have been considered.
\begin{figure*}[!h]
\centering
\includegraphics[width = 0.62\textwidth]{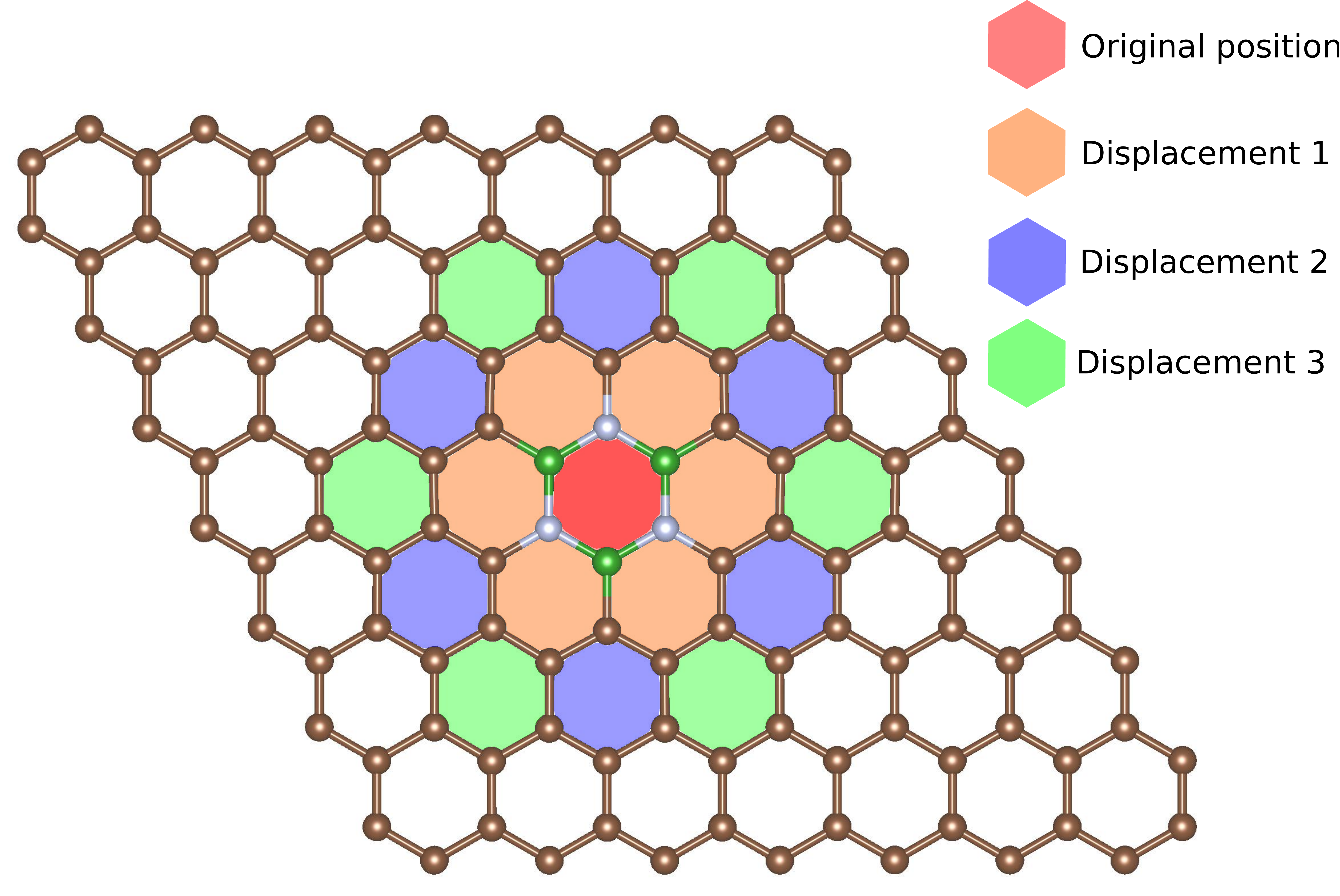}
\caption{Figure S7: Possible displacements of the BN ring to positions around its original one (highlighted in red) in the weakly disordered systems. Single borazine ring (see Fig.1a in the main text) in the original cell is considered. Three disorder models were considered in this work: Model M$_1$ with displaced BN rings in the orange positions; Model M$_2$ with orange and blue positions; Model M$_3$ with displaced rings in all orange, blue and green positions.}
\label{fig_sim7}
\end{figure*}

With the described disorder models, the electronic structure presented in this paper is computed using a simple $p_z$-orbital tight binding Hamiltonian \cite{Fiori12,Nguyen12} with a proper adjustment for B-N hopping energies. The validity of this tight-binding Hamiltonian is illustrated in by a comparison of the computed band structure with the DFT results for periodic systems (see an example in Fig.S8).

\begin{figure*}[!h]
\centering
\includegraphics[width = 0.7\textwidth]{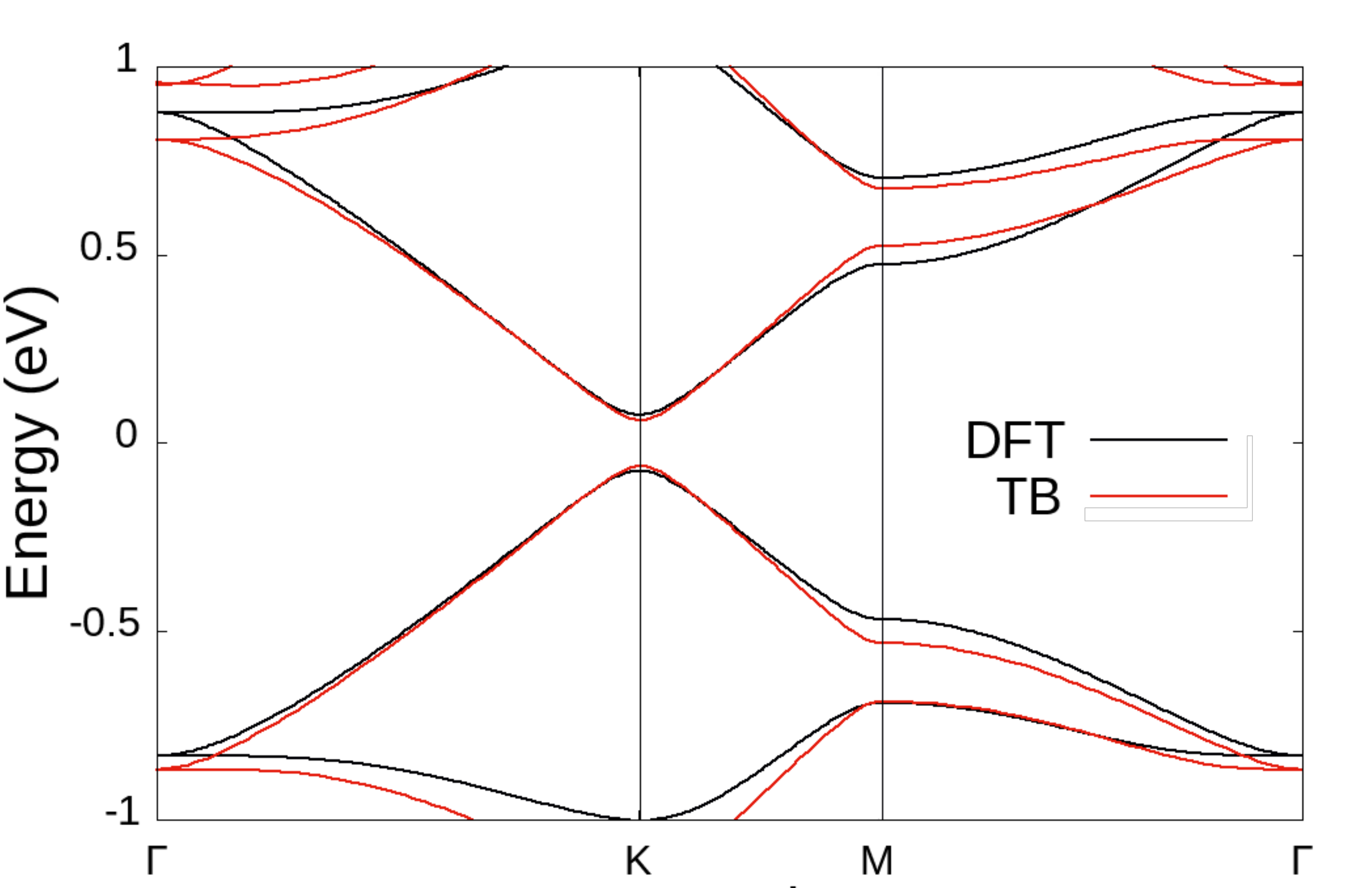}
\caption{Figure S8: Tight-binding bandstructure compared to the DFT results for the periodic system of 8$\times$8 cell containing a single borazine (BN)$_3$ ring (see model 1 in the main text).}
\label{fig_sim8}
\end{figure*}

In addition to the results in Fig.6 of the main text, we present in Fig.S9 the electronic structure of BN-ring doped systems when three different disorder models aforementioned are taken into account. Interestingly, the bandgap is shown to be almost unaffected by the considered location disorders. This implies that graphene sublattice symmetry breaking induced by the presence of BN-rings is the essential ingredient (see the detailed discussions in the main text) whereas the spatial variation of BN-ring distance does not play important roles on the bandgap opening.

\begin{figure*}[!h]
\centering
\includegraphics[width = 0.98\textwidth]{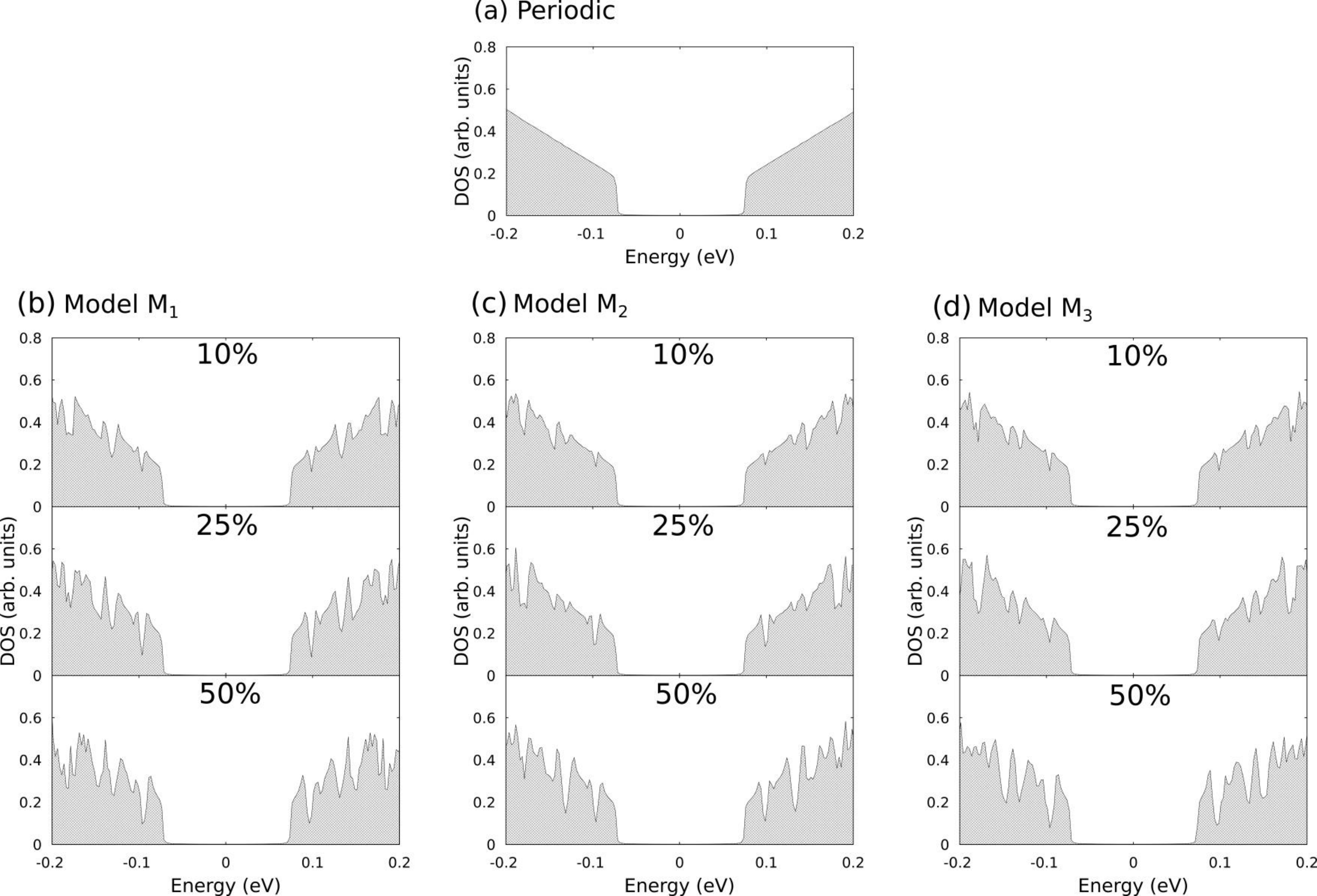}
\caption{Figure S9: Electronic density of states of (a) periodic system compared to the results in disordered ones: (b) disorder Model M$_1$, (c) Model M$_2$ and (d) Model M$_3$ with different disorder probabilities as indicated. All considered systems are created using the original 8$\times$8 cell containing a single borazine ring (see Fig.1a in the main text).}
\label{fig_sim9}
\end{figure*}

\end{document}